\documentclass[prd,superscriptaddress,amsfonts,amssymb,amsmath,showpacs,twocolumn]{revtex4-2}
\usepackage{bm}
\usepackage{amsfonts}
\usepackage{latexsym}
\usepackage{graphicx}
\usepackage{amsmath}
\usepackage{palatino}
\usepackage{mathpazo}
\usepackage{textcomp}
\usepackage{float}
\usepackage{booktabs}
\usepackage{dcolumn}
\usepackage{booktabs}
\usepackage{multirow}
\usepackage{hyperref}
\usepackage{orcidlink}
\hypersetup{colorlinks,citecolor=blue}
\usepackage[caption=false]{subfig}
\usepackage{xcolor}

\begin{document}

\color{black}       

\title{Interacting Ghost Dark Energy with Sign-Changeable Coupling in Brans-Dicke Cosmology}

\author{Kirti Mehta\orcidlink{0009-0001-3149-9890}}
\email{mehtakirti94@gmail.com}
\affiliation{Department of Mathematics, Maharaja Agrasen University, Kalujhanda, Himachal Pradesh-174103, India} 

\author{Pankaj Kumar\orcidlink{0000-0002-7156-5637}}
\email{pankaj.11dtu@gmail.com}
\affiliation{Department of Mathematics, Maharaja Agrasen University, Kalujhanda, Himachal Pradesh-174103, India}

\author{N. Myrzakulov\orcidlink{0000-0001-8691-9939}
}\email{nmyrzakulov@gmail.com}
\affiliation{L N Gumilyov Eurasian National University, Astana 010008, Kazakhstan}

\author{ S. H. Shekh\orcidlink{0000-0002-1932-8431}}\email{  da\_salim@rediff.com}
\affiliation{Department of Mathematics, S.P.M. Science and Gilani Arts, Commerce College, Ghatanji, Yavatmal, \\Maharashtra-445301, India.}
\affiliation{L N Gumilyov Eurasian National University, Astana 010008, Kazakhstan}

\begin{abstract}
\textbf{Abstract:} In this study, we analyze the ghost dark energy model in Brans-Dicke cosmology in the framework of a flat Friedmann-Lemaitre-Robertson-Walker universe. We consider an interaction between ghost dark energy and dark matter with a sign changeable interaction term. To discuss the cosmological implications of the model, we consider a well-motivated logarithmic form of the Brans-Dicke scalar field. By deriving the cosmological evolution equations, we obtain the cosmological parameters such as the equation of state and deceleration parameters. We analyze the behavior of the cosmological parameters by plotting their graphs against the redshift parameter ($z$). We observe that the equation of state parameter shows quintessence-like behaviour during present and future epochs; however, phantom-like behavior is also possible for suitable values of the model parameters. Analysis of the deceleration parameter shows a smooth recent phase transition of the universe (deceleration to acceleration). An interesting result we observe is the decelerated expansion of the universe in the far future, i.e, the universe experiences another phase transition in the future. The physical significance of the well-known cosmological plane ($w_D-w_D'$ plane) is discussed in our model. We observe that the trajectories start in the freezing region with the same initial behavior, deviate from each other during the evolution and ends in the thawing region. Finally, we perform a detailed thermodynamic analysis and demonstrate that the generalized second law of thermodynamics is satisfied within the present interacting ghost dark energy model.

\end{abstract}

\maketitle

\section{Introduction}

Accelerated expansion of the universe has been one of the most important cosmological discoveries in the past few decades. This acceleration was first observed in 1998 through the study of Type Ia supernovae by two independent teams led by Riess and Perlmutter \cite{riess, perl, perlm}. Since then, accelerated expansion has been further confirmed by numerous other observations including the Baryon Acoustic Oscillations \cite{eise}, Cosmic Microwave Background data \cite{kom}, Wilkinson Microwave Anisotropy Probe (WMAP) \cite{sper, hin}, Sloan Digital Sky Survey (SDSS) \cite{teg, tegm} and recently by the observations of the Planck Probe \cite{ade}. This accelerated expansion remains one of the biggest unsolved problems in modern cosmology. Within the framework of general relativity (GR), accelerated expansion implies the existence of an unknown component called as dark energy (DE) which accounts for approximately  $70\%$ of the total energy density of the universe. The most widely accepted model of DE is the $\Lambda$CDM model, where $\Lambda$ is the cosmological constant which is thought to be the energy of quantum vacuum and CDM stands for cold dark matter. Despite having a number of good features, the $\Lambda$CDM model faces serious problems such as the fine-tuning problem, cosmic coincidence problem and age problem \cite{peeb, stei, zlat,roos}. The most recent problem associated with the $\Lambda$CDM model is $H_0$ tension \cite{sah, din, di} which is related to the present value of the Hubble parameter. To develop a more accurate model of the universe, various DE models with time-varying equation of state parameter have been proposed. These models are supported and constrained by observational data \cite{uala, gong, alam}. Examples of such DE models include quintessence \cite{chou, ratr}, phantom \cite{cald, cal}, K-essence \cite{chib, arm}, Chaplygin gas \cite{kam, ben}, holographic dark energy \cite{li, pav, wan, shey, sheyk} and agegraphic dark energy \cite{wei, kim, wu} etc.

A dedicated observational probe for the search of Dark Energy (DE), the DE Spectroscopic Instrument (DESI), indicating towards the dynamic nature of DE \cite{Adame, karim}. In a very resent paper \cite{gu}, the authors analysed measurements of baryon acoustic oscillations in DESI Data and found that the equation of state for DE $w(z)$ varies with redshift, i.e., it is dynamic in nature. Thus, there is growing evidence in favor of dynamic DE. Therefore, it is significant to consider a dynamic DE model to explain the expansion history of the universe. An important candidate for dynamical DE is ghost dark energy (GDE) which can explain DE without entering new degrees of freedom. The GDE is motivated by the idea that quantum gravitational effects at large scales could explain the accelerated expansion of the universe. This model is rooted in the Veneziano ghost field originally proposed to resolve the $U(1)$ problem in Quantum Chromodynamics (QCD) theory \cite{wit, ven, nat}. The $U(1)$ problem emerges from the fact that the QCD Lagrangian exhibits a global chiral $U(1)$ symmetry in the massless limit, however this symmetry does not appear in the spectrum of light pseudoscalar mesons. While the ghost field does not contribution to the vacuum energy in Minkowski spacetime, it produces a non-vanishing contribution in time-dependent or non-flat backgrounds with magnitude proportional to $\Lambda^3_{\text{QCD}}H$, where $H$ denotes the Hubble parameter and $\Lambda_{\text{QCD}}$ represents the QCD mass scale \cite{ohta}. Within the GDE framework, this ghost field vacuum energy can be interpreted as a dynamical cosmological constant \cite{cai,she}. Another important property of the GDE is that it provides the required amount of energy to accelerate the universe. Its energy has the order $\Lambda^3_ {QCD}H$, where $H$ is the Hubble parameter and the QCD mass scale is denoted by $\Lambda^3_{QCD}$. At present time, we have $\Lambda_{QCD}\sim100M_eV$ and $H\sim10^{-33}eV$ which provide $\rho_D\sim(3\times10^{-3}eV)^4$ for GDE \cite{ohta}. This is the correct amount of energy required to explain the accelerated expansion of the universe. Thus the GDE model provides an explanation for the fine-tuning problem associated with the $\Lambda$CDM model. The motivation to consider the GDE is that it is a dynamical DE candidate, requires no additional degrees of freedom and is free from the fine-tuning problem.

Recent investigations have significantly expanded our understanding of GDE across various theoretical frameworks. The model's applicability has been extended beyond flat universe scenarios with comprehensive analyses in non-flat cosmological settings \cite{she, dva, care, def, fen}. Furthermore, researchers have successfully reconstructed the GDE model using different theoretical approaches notably through tachyon and quintessence formulations \cite{shey, sheyk}. Biswas et al. \cite{bis} have discussed the generalized GDE in the Friedmann-Lemaitre-Robertson-Walker (FLRW) framework and constrained the model. They showed that the model was stable and explained the recent accelerated expansion of the universe. In the paper \cite{cruz}, authors discussed GDE model and some of its generalisations. The authors showed that the model exhibits type-III future singularity if the GDE model is extended with $\dot H$ term. Furthermore, this extended model also exhibits instability under the squared sound speed criteria.

The Brans-Dicke (BD) theory provides a dynamical framework for DE models, especially  dynamical DE models, to study cosmic evolution. However, it is a natural extension of general relativity (GR) which accommodates Mach's principle. The GDE model inherently represents a dynamical cosmological constant, and its investigation within the BD theoretical framework emerges as a more natural approach compared to traditional Einstein's gravity. BD theory as a dynamical framework has been considered by a large number of authors to study dynamical DE models \cite{as,cps,gh,ka}. These properties motivated us to discuss the GDE model within the dynamical framework of BD theory. We consider a possible interaction between the GDE and dark matter (DM) and choose a sign-changeable interaction term for our study.  The sign-changeable interaction term $Q$ has been formulated in \cite{we, weH} expressed as $Q = q(\alpha\dot{\rho} + 3\beta H\rho)$ where $\alpha$ and $\beta$ are constants. The interaction term $Q$ exhibits sign reversal behavior that correlates with changes in $q$ during universal evolution. On the other hand, it provides the opportunity for both way energy transfer between DE and DM during different epochs of cosmic evolution. Indeed, this is a significant property and is very possible in the present dark epoch of cosmic evolution. Therefore, it is important to discuss the GDE model by considering the sign-changeable interaction term. Here, we take a sign-changeable interaction term  $Q = 3b^2Hq\rho_D$ to discuss our GDE model within the BD theory.

In this study, we investigate interacting GDE model in the BD theory. We derive the cosmological evolution equations and study the resulting dynamics using the equation of state (EoS) and deceleration parameter (DP) by choosing a sign changeable interaction term for our analysis. We apply the $w_D-w_D'$ diagnostic and thermodynamic analysis to our model. Our goal is to provide a comprehensive analysis of interacting GDE model in the BD theory and assess its viability as an alternative to standard $\Lambda$CDM cosmology.

The structure of this paper is organized as follows. Sections 2 and 3 introduce the field equations of BD theory incorporating the interacting GDE and a logarithmic form of the BD scalar field. In Section 4, we derive the cosmological parameters including the EoS parameter and DP to explore the potential evolution of the universe. In Section 5, we examine the \(w_{D}-w'_{D}\) trajectories for various values of model parameters. Section 6 discusses the thermodynamic analysis of the model and Section 7 concludes the study with a summary of our findings.
\section{Model and Field Equations}
In this section, we present the mathematical framework for BD theory that incorporates the GDE. We begin with the action of the BD theory in the Jordan frame, which includes a matter Lagrangian density $\mathcal{L}_m$ given as
\begin{equation}
	S = \int  \left[\frac{1}{2}\left(-\phi R + \frac{\omega}{\phi}g^{\mu\nu}\partial_\mu\phi\partial_\nu\phi\right) + \mathcal{L}_m\right]\sqrt{-g}d^4x,
\end{equation}
where $R$ represents the Ricci scalar curvature, $g$ denotes the determinant of the metric tensor $g_{\mu\nu}$ and $\omega$ is a dimensionless BD coupling parameter. The BD scalar field $\phi$, which was introduced to represent the variable gravitational constant, was defined as $\phi = (8\pi G)^{-1}$.

To discuss our model, we consider a flat FLRW universe given by the line element
\begin{equation}
	ds^{2}=dt^{2}-a^{2}(t)(dx^{2}+dy^{2}+dz^{2}),
\end{equation}
where $a(t)$ representing the cosmic scale factor. The rest of the notations have their usual meaning. Observational evidence also indicates that the universe possesses nearly flat spatial geometry. Our focus in the present study is on late-time cosmic evolution; therefore, we consider a spatially flat universe. As the late universe is dominated by dark sector, therefore, we consider that the universe consists of a GDE and dark matter (DM). Since the present analysis focuses on late-time cosmic evolution dominated by the dark sector, the contribution of baryonic matter is neglected without loss of generality.

The field equations derived from the BD action and FLRW metric are
\begin{equation}
	H^2+H\frac{\dot\phi}{\phi}-\frac{\omega}{6}\frac{\dot\phi\;^2}{\phi\;^2}=\frac{\rho_{m}+\rho_{D}}{3\phi},
\end{equation}
\begin{equation}
	2\frac{\ddot a }{a}+H^2+2 H\frac{\dot\phi}{\phi}+\frac{\omega}{2}\frac{\dot\phi\;^2} {\phi\;^2}+\frac{\ddot\phi}{\phi}=-\frac{p_{D}}{\phi},
\end{equation}
\begin{equation}
	\ddot\phi+3H\dot\phi=\frac{\rho_{m}+\rho_{D}-3p_{D}}{2\omega+3},
\end{equation}
Here, $H = \dot{a}/a$ is the Hubble parameter, and $\rho_m$ and $\rho_D$ represent the energy densities of the DM and the GDE respectively. The energy density of the GDE is expressed as \cite{ohta}
\begin{equation}
	\rho_D = \gamma H,
\end{equation}
where $\gamma$ is a positive constant.

We adopt the logarithmic form of the BD scalar field proposed in \cite{pankaj}. The well-motivated form is expressed as
\begin{equation}
	\phi = \phi_0 \ln(\alpha + \beta a),
\end{equation}
where $\phi_0$, $\alpha$ and $\beta$ are constants with the conditions $\phi_0>0$, $\alpha > 1$ and $\beta > 0$. These conditions are required to prevent $\phi$ from having a  negative value. This form of the BD scalar field is able can evolve sufficiently slowly to achieve the required slow variation in the BD scalar field. Therefore, the logarithmic form is a viable choice for the BD scalar field.

This formalism provides the foundation for subsequent analysis of the GDE model within the BD theory framework.

\section{Interacting GDE Model with Sign-Changeable Interaction}

The potential interaction between DM and DE remains a significant area of investigation in modern cosmology. Although conclusive evidence for coupling between these dark sectors remains undetermined, specific astronomical observations \cite{cao, cos} suggest the possibility of such an interaction. Numerous theoretical frameworks examining DM-DE coupling have been extensively investigated in the literature \cite{ame, zim, farr, gum, sko}. A recent study \cite{pin} examined the coupling between new agegraphic dark energy and DM in BD cosmology focusing on sign-changeable interaction terms in flat spacetime geometry. Taking advantage of observational support for interaction, we consider the interaction between the DM and the GDE. The continuity equations in the presence of the interaction between the DM and the GDE can be written as:
\begin{align}
	\dot{\rho}_m + 3H\rho_m &= Q, \\
	\dot{\rho}_D + 3H(1 + w_D)\rho_D &= -Q,
\end{align}
where $w_D = p_D/\rho_D$ is the EoS parameter of the GDE and $Q$ is the interaction term. The sign of the interaction term $Q$ governs the direction of energy transfer between the DM and the GDE, where $Q > 0$ signifies the energy flow from the GDE to the DM and $Q < 0$ indicates the reverse process. Observational data indicates sign reversal in dark sector interaction within $0.45 \leq z \leq 0.9$ \cite{cai}. Wei \cite{wei, xu} addressed this by proposing a sign-changeable interaction term $Q = q(\alpha\dot{\rho} + 3\beta H\rho)$, where $\alpha$, $\beta$ are constants and $q$ denotes the DP. This formulation enables $Q$ to change its sign as $q$ varies during cosmic evolution. We examine the interaction between DM and GDE in BD theory by utilizing a sign-changeable term with a logarithmic BD scalar field.
We adopt a simple sign-changeable interaction term of the form
\begin{equation}
	Q = 3b^2Hq\rho_D,
\end{equation}
where $b^2$ is the coupling constant and $q$ is the DP.

\section{Evolution of the Universe}
Various cosmological parameters are essential to understand the evolution of the universe. Among these, the DP $(q)$ and EoS parameter $(w_D)$ hold significant importance. We discuss the evolution of the universe through the EoS parameter and DP for our model.

The EoS parameter is one of the important cosmological parameters used to determine the features of accelerated expansion of the universe. It is defined as $w_D=p_D/\rho_D$. The value of $w_D$ is categorized as follows: $w_D=1$, it describes a stiff fluid, $w_D=-1$ represents the cosmological constant, $w_D=1/3$ describes radiation dominated era $w_D=0$ represents a matter dominated phase, $w_D<-1$ describes a phantom type DE dominated era and  $-1<w_D<-1/3$ represents a quintessence type DE dominated era. \\
Using Eqs. (6), (7) and (10) in (9), we obtain the value of the EoS parameter for the GDE which is given by
\begin{equation}
	w_D = \frac{-2}{3} + \frac{q (1- 3b^2)}{3},
\end{equation}
where $q$ is the DP defined by $q = -\frac{\ddot{a}}{a H^2}= -1 - \frac{\dot{H}}{H^2}.$ Equation (13) shows that the equation of state parameter explicitly depends on the deceleration parameter and the interaction strength. Consequently, any phase transition in cosmic expansion directly influences the DE dynamics. It is evident from the expression for  $w_D$ given by Eq.~(13) to explain the evolution of $w_D$ we require $q$. Additionally, $q$ plays a crucial role in describing the expansion of the universe. Its sign indicates whether the universe is undergoing acceleration or deceleration or not. If the value of DP is negative ($q<0$) then the universe undergoes accelerating expansion, for a positive value ($q>0$) it exhibits decelerating expansion and a change in sign represents the phase transition of the universe. The universe has experienced two fundamental phase transitions across its evolutionary time line, one during its earliest epoch and the other in the recent cosmological past. The value of $q$ can be obtained using Eqs. (6) and (7) in Eq. (4) as follows
\begin{widetext}
\begin{equation}
	q = \frac{1 + \frac{2 \beta a}{(\alpha + \beta a)\ln(\alpha + \beta a)} - \frac{\beta^2 a^2}{(\alpha + \beta a)^2 \ln(\alpha + \beta a)} + \frac{\omega \beta^2 a^2}{2(\alpha + \beta a)^2 [\ln(\alpha + \beta a)]^2}+\frac{w_{D} \gamma}{ H \phi_0 \ln(\alpha + \beta a)}}{2 + \frac{\beta a}{(\alpha + \beta a)\ln(\alpha + \beta a)}}
\end{equation}

\end{widetext}
Here, it is observed that the expression for $q$ depends on $w_D$, as $w_D$ also has $q$ in the expression. Overall, Fig. 1 demonstrates that the interacting GDE model supports quintessence and phantom regimes while remaining consistent with observational bounds on the present equation of state. To determine these parameters independently, we utilize Eqs.~(13) and~(14) which contain both $w_D$ and $q$. By solving these equations simultaneously, we derive the values for $w_D$ and $q$. The final value of $w_D$ is given by
\begin{widetext}
\begin{equation}
	w_D = -\frac{2}{3} + \frac{(1-3b^2)}{3}\left[\frac{1 + \frac{2 \beta a}{(\alpha + \beta a)\ln(\alpha + \beta a)} - \frac{\beta^2 a^2}{(\alpha + \beta a)^2 \ln(\alpha + \beta a)} + \frac{\omega \beta^2 a^2}{2(\alpha + \beta a)^2 [\ln(\alpha + \beta a)]^2}-\frac{2 \gamma}{3 H \phi_0 \ln(\alpha + \beta a)}}{2 + \frac{\beta a}{(\alpha + \beta a)\ln(\alpha + \beta a)}- \frac{\gamma(1-3b^2)}{3 H \phi_0 \ln(\alpha + \beta a)}}\right]
\end{equation}
\end{widetext}

\begin{figure*}
	\subfloat[]{\includegraphics[width = 8cm,height=5.2cm]{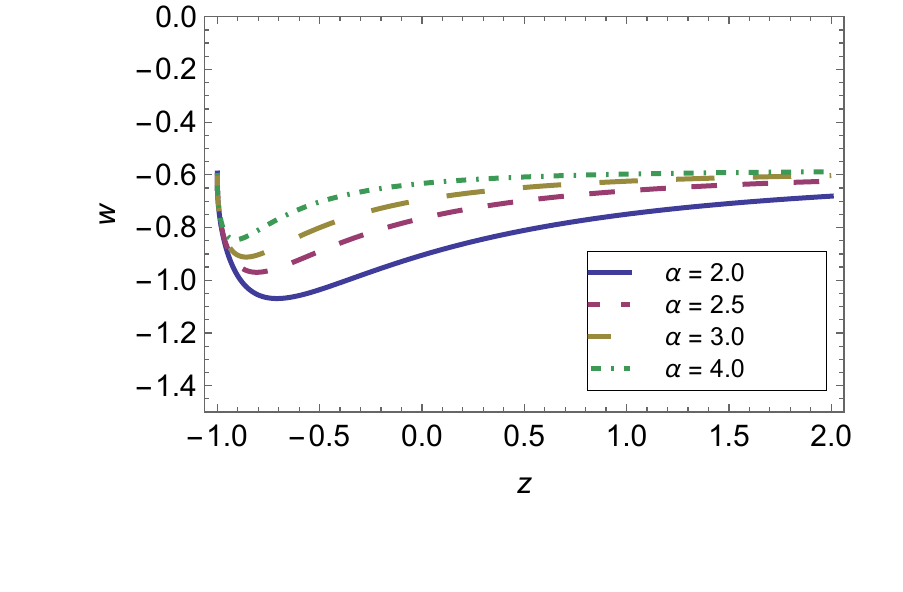}}\;
	\subfloat[]{\includegraphics[width = 8cm,height=5.2cm]{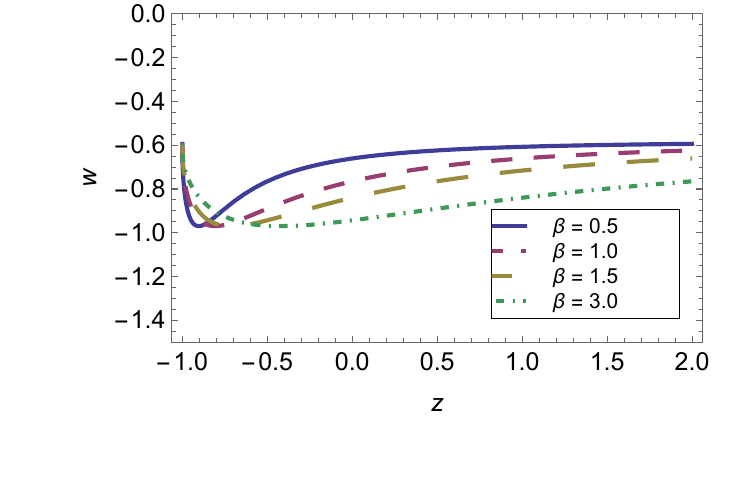}}\;
	\subfloat[]{\includegraphics[width = 8cm,height=5.2cm]{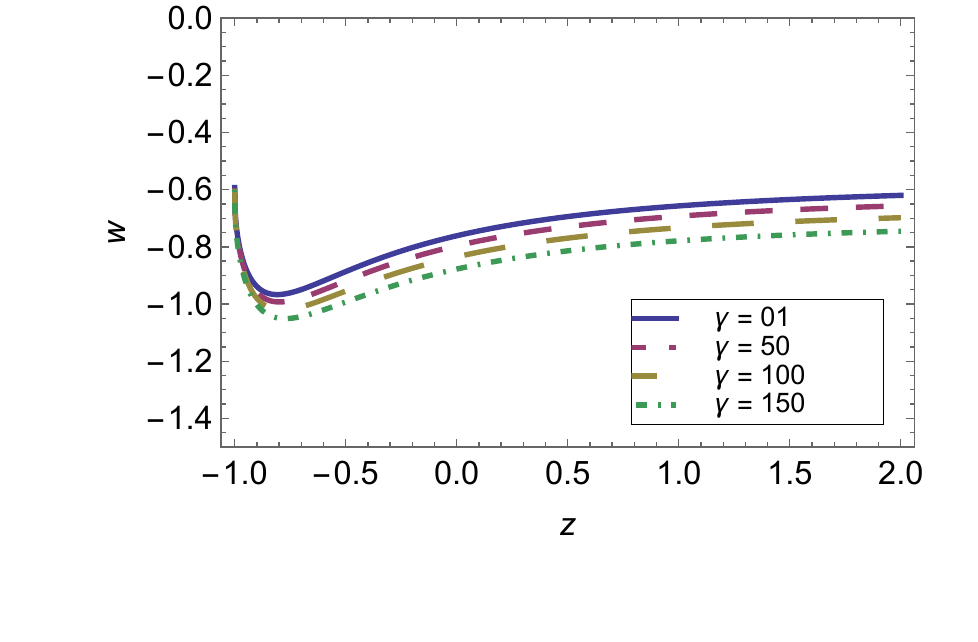}}\;
	\subfloat[]{\includegraphics[width = 8cm,height=5.2cm]{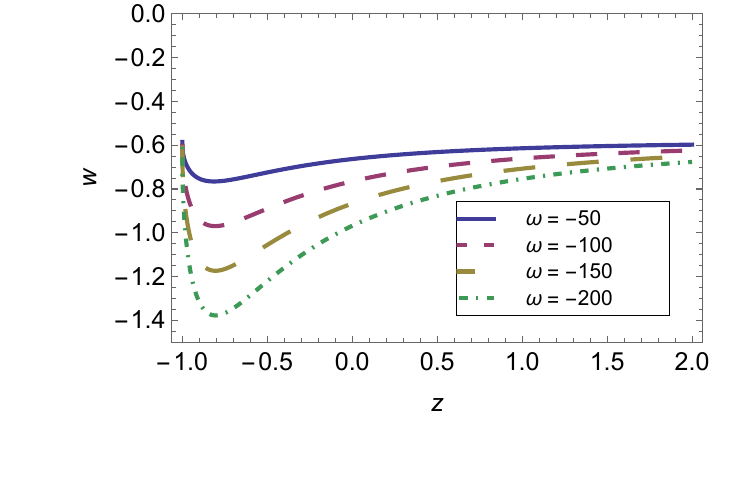}}
	\caption{We have taken $H=70$, $\phi_0=1$, $b=0.4$ and $\gamma=6$ to plot $w_D$ against $z$. Fig.(a) is plotted for fixed values $\omega=-100$, $\beta=1$ and for various values of $\alpha$. Fig.(b) is plotted for fixed values $\omega=-100$, $\alpha=2.5$ and for various values of $\beta$. Fig.(c) is plotted for fixed values $\alpha=2.5$, $\beta=1$ and for various values of $\omega$. Fig.(d) is plotted for fixed values $\alpha=2.5$, $\beta=1$, $\omega= -100$ and for various values of $\gamma$.}
\end{figure*}

Let us discuss the evolution of the universe through the EoS parameter ($w_D$) of the GDE. It is more applicable to discuss the growth of the universe using cosmological parameters as functions of $z$ rather than $a$. Therefore, using the relation  $a=(1+z)^{-1}$ in normalized unit, we have plotted graphs of $w_D$. The plots of $w_D$ against $z$ are shown in Fig.~1(a-d) for different values of the model parameters. We vary one model parameter of $\alpha$, $\beta$, $\omega$, and $\gamma$, while keeping the other parameters fixed in a sub-figure, like we vary $\alpha$ and take fixed values of others in Fig. 1(a). It is evident from the evolution trajectories of $w_D$ in the Fig. (1) that EoS of DE has been achieved which is required to explain the present accelerated expansion of the universe. Notably, the phantom divide line (PDL) ($w_D = -1$) can be crossed during the present evolution of the universe. As illustrated in Fig.~1(a), the PDL is crossed for smaller values of the parameter $\alpha$. It is worth noting that the model shows the weakening of GDE in the future. This result is in agreement with recent results of the DESI survey \cite{Adame,karim} which predicts the dynamic nature of DE and weakening of DE over the time. Thus, the model showing agreement with the latest observations.

A recent study \cite{brou} determined the present value of EoS parameter for DE to be $w_0 = -0.90 \pm 0.14$ based on SNIa observations alone and $w_0 = -0.978^{+0.024}_{-0.031}$ from a combined analysis of SNIa and Cepheid variables using data from the SH0ES collaboration. Meanwhile, according to \cite{agh}, $w_0$ was measured as $-1.03 \pm 0.03$. The $\Lambda$CDM model produces a constant EoS parameter $w = -1$. The dynamical analysis of our interacting GDE model demonstrates consistency with observational constraints on $w_0$ for appropriate model parameters (Fig.~1). Moreover, in the present model $w_0$ falls within the range $-0.95 \leq w_0 \leq -0.65$ for most of the parameter combinations. However, phantom-crossing behavior may be achieved for suitable parameter combinations as discussed above. Our findings indicate that this model provides a viable framework for describing DE evolution offering rich dynamics while maintaining consistency with the observational constraints.

The final expression for $q$ is obtained by solving Eqs. (13) and (14) is given as
\begin{widetext}
\begin{equation}
	q = \frac{1 + \frac{2 \beta a}{(\alpha + \beta a)\ln(\alpha + \beta a)} - \frac{\beta^2 a^2}{(\alpha + \beta a)^2 \ln(\alpha + \beta a)} + \frac{\omega \beta^2 a^2}{2(\alpha + \beta a)^2 [\ln(\alpha + \beta a)]^2}-\frac{2 \gamma}{3 H \phi_0 \ln(\alpha + \beta a)}}{2 + \frac{\beta a}{(\alpha + \beta a)\ln(\alpha + \beta a)}- \frac{\gamma(1-3b^2)}{3 H \phi_0 \ln(\alpha + \beta a)}}
\end{equation}
\end{widetext}

\begin{figure*}
	\subfloat[]{\includegraphics[width = 8cm,height=5.2cm]{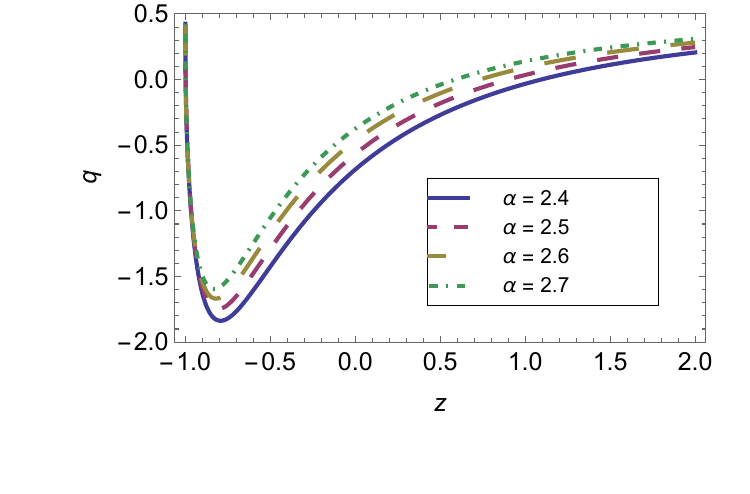}}\;
	\subfloat[]{\includegraphics[width = 8cm,height=5.2cm]{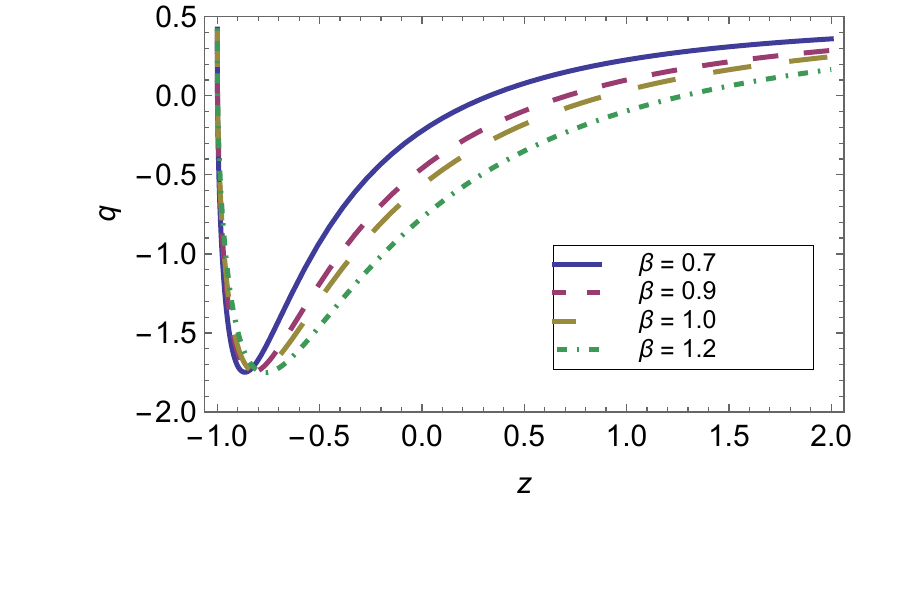}}\;
	\subfloat[]{\includegraphics[width = 8cm,height=5.2cm]{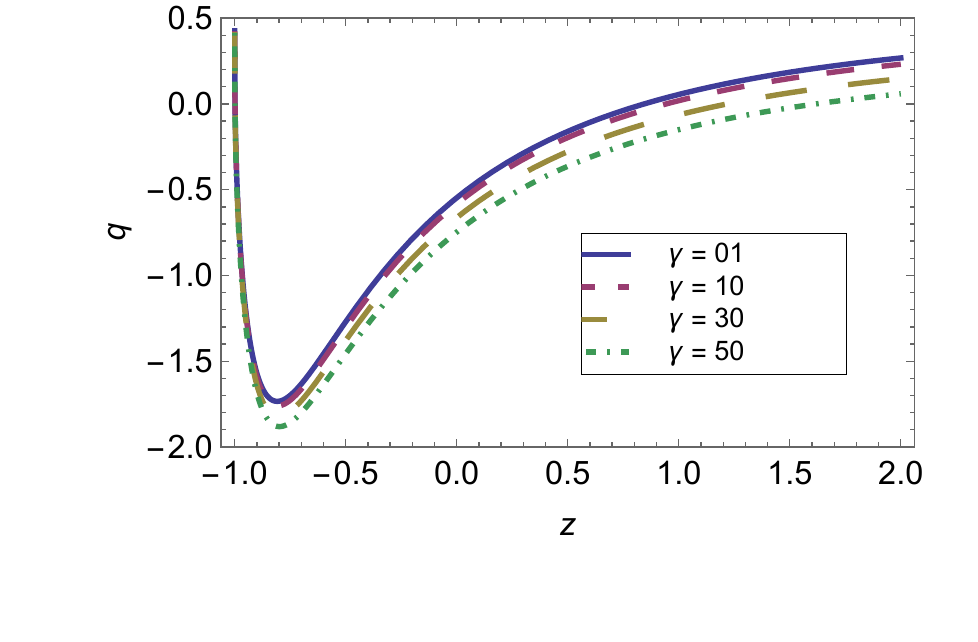}}\;
	\subfloat[]{\includegraphics[width = 8cm,height=5.2cm]{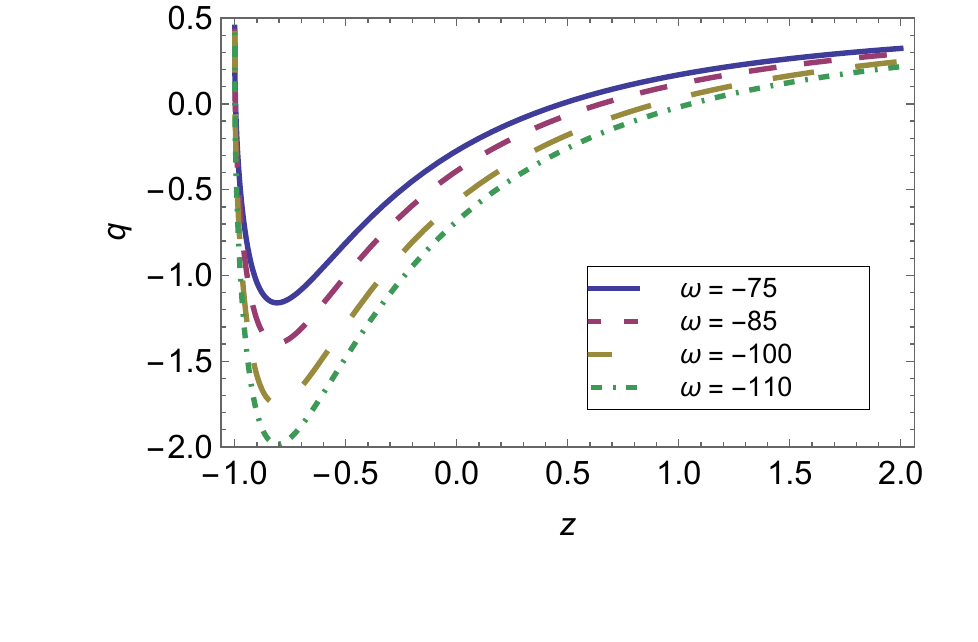}}
	\caption{We have taken $H=70$, $\phi_0=1$ and $b=0.4$ to plot $q$ against $z$. Fig.(a) is plotted for fixed values $\omega=-100$, $\beta=1$, $\gamma=6$ and for various values of $\alpha$. Fig.(b) is plotted for fixed values $\omega=-100$, $\alpha=2.5$, $\gamma=6$ and for various values of $\beta$. Fig.(c) is plotted for fixed values $\alpha=2.5$, $\beta=1$, $\gamma=6$  and for various values of $\omega$. Fig.(d) is plotted for fixed values $\alpha=2.5$, $\beta=1$, $\omega= -100$ and for various values of $\gamma$.}
\end{figure*}

Now, we analyze the evolution of the universe through $q$. Using the relation $a=(1+z)^{-1}$ in the normalized unit, we have plotted the trajectories of $q$ versus $z$ in Fig.~2. It is evident from the figure that the model shows a smooth recent phase transition of the universe. Moreover, it is worthwhile to note that it shows a decelerated expansion in the future which is confirms the weakening of the GDE as shown in the discussion on $w_D$. Thus, the results obtained from the analysis of $q$ also showing agreement with DESI survey results.

According to \cite{lu}, the transition redshift marking the shift from a decelerated to an accelerated phase in the $\Lambda$CDM model is estimated as $z_t = 0.761 \pm 0.055$. Our analysis, as illustrated in Fig.~2(a-d), shows that for most parameter combinations in the present model, the transition redshift lies within the range $0.3 < z_t < 1.6$, demonstrating consistency with observational estimates. Recent measurements \cite{rie} report the present value of the DP as $q_0 = -0.51 \pm 0.024$, while Capozziello et al. \cite{capo} obtained $q_0 = -0.56^{+0.04}_{-0.04}$. The evolution of $q(z)$ depicted in Fig.~2 indicates that, for suitable choices of model parameters, the present value of $q$ lies within the observational bounds \cite{rie, capo}, i.e., $-0.8 < q_0 < -0.2$. All trajectories of $q(z)$ consistently exhibit a transition from an early decelerated to a late-time accelerated phase, which is in agreement with observational evidence. The influence of the model parameters $\alpha$, $\beta$, $\omega$, and $\gamma$ is shown in Figs. 2(a), 2(b), 2(c), and 2(d), respectively. For instance, smaller values of $\alpha$ lead to an earlier (higher-redshift) transition from deceleration to acceleration. Overall, the model provides a viable description of late-time cosmic acceleration, predicts a possible future decelerated phase, and remains consistent with the current cosmological observations. The emergence of a future decelerated phase may help to improve the eternal acceleration problem, and has important consequences for the ultimate fate of the universe.

\section{The $w_D-w_D'$ analysis}
The $w_D-w_D'$ diagnostic method was introduced by Caldwell and Linder \cite{lin}. Investigation of the trajectories of DE models in $w_{D}-w_{D}'$ plane provides a technique for classification of the model using the model parameters. Here, $w_D'$ denotes the derivative of $w_D$ with respect to the logarithm of the scale factor, i.e., $w_D'=dw_D/d\ln a$. Each model maps out unique trajectories allowing to distinguish and classify DE models. The $\Lambda$CDM model has fixed points $w_D=-1$, $w'_D=0$ in $w_{D}-w_{D}'$ plane. In the original paper, the authors characterized two distinct classes known as the thawing region ($w_{D}'>0$) and freezing region ($w_{D}'<0$) on the $w_D-w_{D}'$ plane.

We apply the analysis to study the behavior of the interacting GDE model in the framework of BD theory. We have plotted the trajectories in $w_{D}-w_{D}'$ plane to observe the behavior the model in Fig. 3. It is observed that the trajectories lie in the freezing region during the present epoch characterized by negative values of $w_D'$. The analysis reveal that trajectories in the $w_{D}-w_{D}'$ plane begin to show a thawing region during  future evolution. It can be clearly observed from the figure that the trajectories start in the freezing region with the same initial behavior, deviate from each other during the evolution and end in the thawing region showing the same behavior in the future. The evolution pattern of these trajectories in the $w_{D}-w_{D}'$ plane appears to be dependent on the model parameters. Notably, the trajectories may pass through the $\Lambda$CDM model ($w_D=-1$, $w_D'=0$) for suitable values of the model parameters during the evolution. The trajectories show a unique evolutionary behavior for our model of interacting GDE in the BD theory. Unlike the $\Lambda$CDM model, which corresponds to a fixed point in the \( w_D - w_D' \) plane, the present model exhibits evolving trajectories, highlighting its genuinely dynamic nature. This transition indicates a gradual departure from slow-roll DE behavior toward a more dynamic regime in the future.
\begin{figure}
	\begin{center}
		\includegraphics[width=9cm, height=7cm]{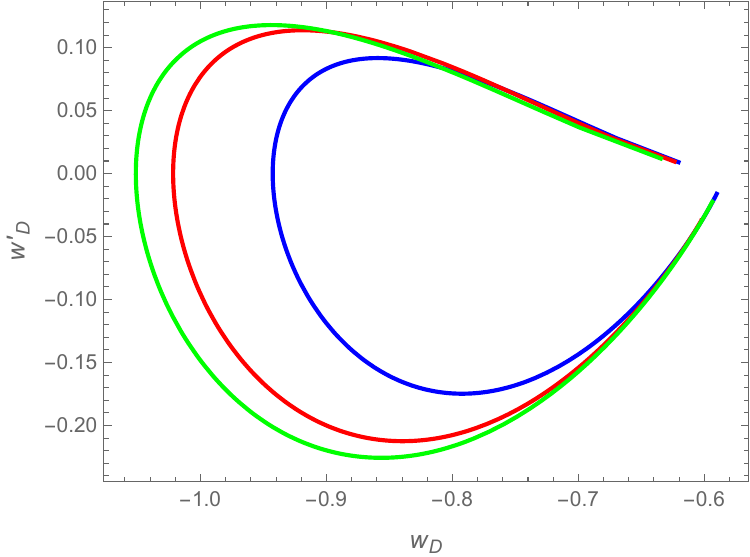}
		\caption{We have plotted trajectories in $w_{D}-w_{D}'$ plane and have taken fixed values of $\phi_0= 1$, $H= 70$, $\gamma= 6$, $b= 0.4$. The trajectories are plotted for various values of parameters $\alpha$, $\beta$ and $\omega$. The trajectory 1 has been plotted for $\alpha = 2.4$, $\beta = 1.5$, $\omega = -90$. We have taken $\alpha = 2.2$, $\beta = 1.8$, $\omega = -100$ to plot trajectory 2. The trajectory 3 has been plotted for $\alpha = 2.5$, $\beta = 1.6$, $\omega = -120$.}
	\end{center}
\end{figure}
\section{Thermodynamic Analysis}
The connection between gravity and thermodynamics has been the subject of significant interest in cosmology. Jacobson \cite{jac} derived Einstein's field equations from thermodynamic considerations while Padmanabhan \cite{padm} showed that the first law of thermodynamics can be obtained from Einstein's field equations for static, spherically symmetric space-time. These study show a direct connection between the two theories and provide a framework for thermodynamic studies in gravitation theories. The generalized second law of thermodynamics (GSL) is an important part of thermodynamic analysis and has been discussed in both Einstein's theory and modified theories of gravity \cite{ mom, mam, kar, duar}.

According to GSL, the total entropy of the universe (the sum of the horizon entropy and the entropy of matter inside the horizon) is an increasing function of time. The GSL has been discussed in pure and modified BD theories in papers \cite{bha, dua} respectively. In the present work, we have considered the original BD theory in a flat FLRW universe. Let us now study the thermodynamic implications of the proposed  model using the GSL. The total entropy of the universe is given by
\begin{equation}
	S_{\text{tot}} = S_{\text{h}} + S_{\text{in}},
\end{equation}
where $S_{\text{tot}}$ denotes the total entropy, $S_{\text{h}}$ is the horizon entropy and $S_{\text{in}}$ is the entropy of the matter inside the horizon.
In accordance with thermodynamic principles and the GSL, assuming that the universe is an isolated system, the following entropy inequality must be satisfied
\begin{equation}
	\dot S_{tot} \geq 0.
\end{equation}

The rate of change of total entropy can be expressed as
\begin{equation}
	\dot{S}_{\text{tot}} = \dot{S}_{\text{h}} + \dot{S}_{\text{in}},
\end{equation}
where the over-dot denotes the time derivative. The analysis of the dynamical apparent horizon's entropy presents a more theoretically compelling framework than to the teleological event horizon.
The apparent horizon entropy is given by the following relation \cite{bak}
\begin{equation}
	S_{h} = 2\pi A,
\end{equation}
where $A = 4\pi R_h^2$ denotes the apparent horizon area. In a flat FRW universe, the horizon radius $R_h$ is related to the Hubble parameter as $R_h = \frac{1}{H}$.
Consequently, the apparent horizon entropy emerges as $S_h = \frac{8\pi^2}{H^2}$ and the rate of change of the horizon entropy is given by
\begin{equation}
	\dot{S}_{\text{h}} = -16\pi^2 \frac{\dot{H}}{H^3}.
\end{equation}

By applying Gibb's law of thermodynamics to the fluid enclosed within the horizon, we establish a well-defined relationship expressed as
\begin{equation}
	T_{in}d S_{in} = d E_{in}+ p_{t}d V_{h},
\end{equation}
where the volume $V_{h} = \frac{4}{3}\pi R_{h}^3$. This leads to the rate of change in the entropy of the fluid inside the horizon which is given by
\begin{equation}
	\dot S_{in} = \frac{(\rho_{t}+ p_{t})\dot V_{h}+\dot\rho_{t}V_{h}}{T_{in}}.
\end{equation}
Assuming thermal equilibrium between the fluid inside the horizon and the horizon itself, the temperature inside the horizon ($T_{in}$) is equated to the temperature of the dynamic apparent horizon, also known as the Hayward-Kodama temperature \cite{dua}. This temperature can be mathematically defined as
\begin{equation}
	T_{h} = \frac{2H^2 + \dot H}{4\pi H}.
\end{equation}
It can be observed that this temperature reduces to the Hawking temperature
$T_{\text{Haw}} = \frac{H}{2\pi}$~ \cite{haw} in the de-Sitter space where $\dot{H} = 0$.
Now, the rate of change of the entropy of the fluid inside the horizon can be obtained using Eqs.~(23) and~(24) as follows
\begin{equation}
	\dot{S}_{\text{in}} = 16\pi^2 \frac{\dot{H}}{H^3} \left(1 + \frac{\dot{H}}{2H^2 + \dot{H}}\right).
\end{equation}
Now using Eqs. (19), (21) and (25), the rate of change in the total entropy can be written as
\begin{equation}
	\dot S_{tot} = \frac{16 \pi^2(\frac{\dot H}{H^2})^2}{H(\frac{\dot H}{H^2}+2)}.
\end{equation}
It is clear that $\dot{S}_{\text{tot}}$ depends on the $H$ and its time derivative. We now require the value of $\frac{\dot{H}}{H^2}$ to observe the evolution of $\dot S_{tot}$ and  plot its graphs. The value of $\frac{\dot{H}}{H^2}$ has been calculated by substituting the value of $q$ from equation (16) into the relation $\frac{\dot{H}}{H^2} = -1 - q$. It provides the value as
\begin{widetext}
\begin{equation}
	\frac{\dot{H}}{H^2}= -3-\frac{3\beta a}{(\alpha + \beta a)\ln(\alpha + \beta a)}+\frac{\gamma(1-b^2)}{ H \phi_0 \ln(\alpha + \beta a)}+\frac{\beta^2 a^2}{(\alpha + \beta a)^2 \ln(\alpha + \beta a)}-\frac{\omega \beta^2 a^2}{2(\alpha + \beta a)^2 [\ln(\alpha + \beta a)]^2}
\end{equation}
\end{widetext}
Using the value of $\frac{\dot{H}}{H^2}$, the value of $\dot{S}_{\text{tot}}$ can be written as
\begin{widetext}
\begin{eqnarray}
	\dot{S}_{\text{tot}} &=& \frac{16\pi^2 \left( -3 - \frac{3\beta a}{(\alpha + \beta a)\ln(\alpha + \beta a)} + \frac{\gamma(1-b^2)}{ H \phi_0 \ln(\alpha + \beta a)} + \frac{\beta^2 a^2}{(\alpha + \beta a)^2 \ln(\alpha + \beta a)} - \frac{\omega \beta^2 a^2}{2(\alpha + \beta a)^2 [\ln(\alpha + \beta a)]^2} \right)^{2} }
	{ H \left( -1 - \frac{3\beta a}{(\alpha + \beta a)\ln(\alpha + \beta a)} + \frac{\gamma(1-b^2)}{ H \phi_0 \ln(\alpha + \beta a)} + \frac{\beta^2 a^2}{(\alpha + \beta a)^2 \ln(\alpha + \beta a)} - \frac{\omega \beta^2 a^2}{2(\alpha + \beta a)^2 [\ln(\alpha + \beta a)]^2} \right) }
	\hspace{1cm}
\end{eqnarray}
\end{widetext}
\begin{figure}
	\begin{center}
		\includegraphics[width=10cm, height=7cm]{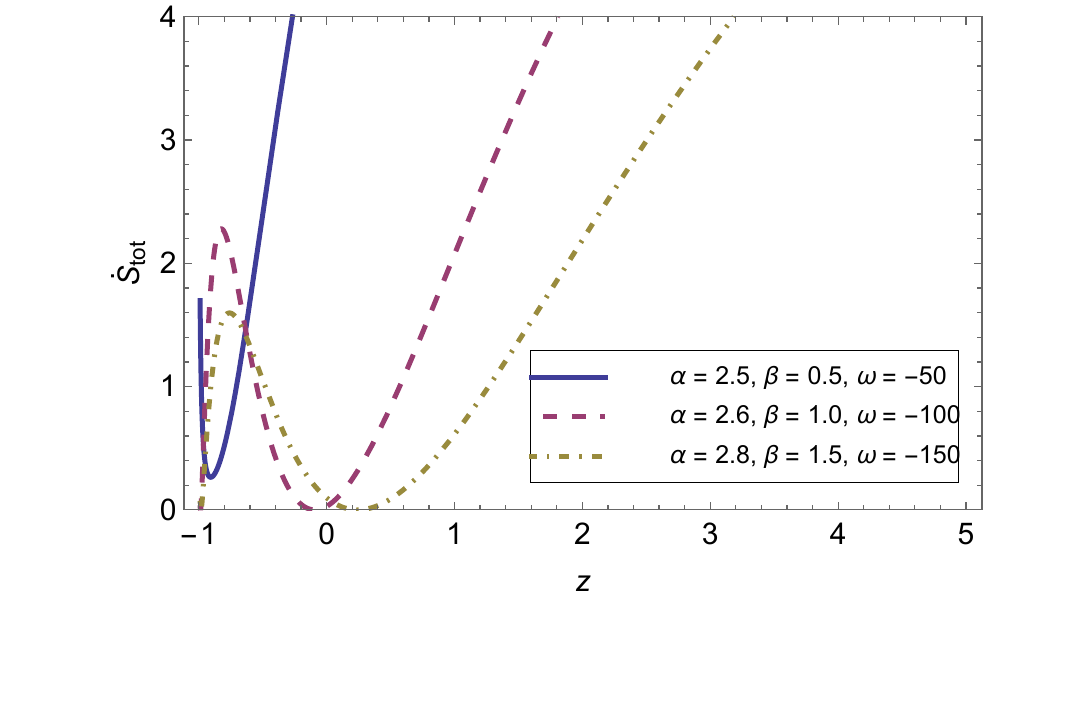}
		\caption{To plot the $\dot{S}_{\text{tot}}$ trajectories, we have taken $H=70$, $\phi_0=1$, $b=0.4$ and $\gamma=6$. The trajectories are plotted for various values of parameters $\alpha$, $\beta$ and $\omega$.}
	\end{center}
\end{figure}
GSL is satisfied if $\dot{S}_{\text{tot}} \geq 0$, which implies $\dot{H}/H^2 \geq -2$. The behavior of $\dot{S}_{\text{tot}}$ as a function of $z$ is plotted in Fig. 4. It is easy to observe from the figure that all trajectories lie in the positive region of $\dot{S}_{\text{tot}}$ which is required to satisfy the GSL. We observe constant total entropy at some points during the present epoch and in the future evolution, i.e., we observe $\dot{S}_{\text{tot}}=0$. The behavior of the trajectories indicates complex thermodynamic interactions within the system while maintaining stability across all temperature regimes and the GSL is indeed satisfied during the present and future evolution of the universe for various combinations of the model parameters. The validity of the generalized second law confirms the thermodynamic consistency and physical viability of the interacting GDE model in Brans--Dicke gravity. Hence, the apparent horizon is preferred over the event horizon since it is locally defined and better suited for dynamic spacetimes.
\section{Concluding Remarks}
In this study, we have undertaken a detailed investigation of the GDE model within the framework of BD gravity, considering a sign-changeable interaction between the DE and DM sectors in a spatially flat FLRW background. The model successfully reproduces the transition from a decelerated to an accelerated phase, which is consistent with current cosmological observations. The DP exhibits the expected phase transition behavior, with the transition redshift being governed by the parameters $\alpha$, $\beta$, $\omega$, and $\gamma$. Moreover, the analysis indicates a possible future decelerating phase, implying that the universe may evolve from its current acceleration towards a later deceleration epoch.

The EoS parameter displays a rich dynamical evolution, smoothly transitioning from a matter-dominated regime to a dark energy-dominated epoch, and allows phantom-divide crossing for appropriate choices of the model parameters. The $w_{D}$--$w'_{D}$ plane analysis reveals a thawing-type behavior during future evolution, clearly distinguishing the present scenario from the standard $\Lambda$CDM cosmology while remaining consistent with observational constraints. A thermodynamic consistency check further confirmed that the model satisfies the GSL throughout the current and future epochs, reinforcing its theoretical robustness. Overall, the interacting GDE scenario in BD gravity provides a viable and self-consistent alternative to the $\Lambda$CDM paradigm, offering a promising framework to account for the late-time acceleration and its possible future evolution. Although confronting the cosmological models with observational datasets is essential for assessing their viability, a full statistical parameter estimation using a cosmic chronometer, baryon acoustic oscillation, and Type Ia supernova data is not implemented in the present work. 
The presence of a logarithmic Brans-Dicke scalar field and a sign-changeable dark sector interaction introduces strong nonlinearities and dynamical feedback that significantly complicate the construction of a stable likelihood function for standard Monte Carlo analyses. Consequently, the present study focuses on a theoretical and phenomenological investigation, ensuring consistency with observationally inferred bounds on key cosmological quantities such as the present values of the equation of state parameter (resembles with DESI data), the deceleration parameter, and the transition redshift. A dedicated numerical framework enabling a direct comparison with observational datasets, including cosmic chronometers, baryon acoustic oscillations, and the Pantheon+ Type Ia supernova compilation, will be developed in future work to further constrain the model parameters and assess the observational viability of the scenario. 

\section*{Financial interests:} The authors have no relevant financial or non-financial interests to disclose. 
\section*{Competing interests:} The authors have no competing interests to declare that are relevant to the content of this article.
\section*{Data availability:} All the data used in this article either has been cited or included in the article itself. No separate data is available for this article.
\section*{Acknowledgments}
The authors N. Myrzakulov and S. H. Shekh thankful to the Science Committee of the Ministry of Education and Science of the Republic of Kazakhstan provided funding for this study (Grant No. AP09058240). The authors (S. H. Shekh and Pankaj Kumar) appreciate the help and resources provided by the IUCAA in Pune, India.

\end{document}